\documentclass[iop]{emulateapj}

\usepackage{color,graphicx}
\usepackage{amsmath,amssymb}
\usepackage{tabularx}

\usepackage[tight]{subfigure}
\usepackage[colorlinks]{hyperref}
\hypersetup{ colorlinks, linkcolor=blue, citecolor=blue, anchorcolor=blue }

\begin{document}


\title{Pre-explosion companion stars in Type I\lowercase{ax} supernovae}

\author{Zheng-Wei Liu\altaffilmark{1},  Richard J. Stancliffe\altaffilmark{1}, C. Abate\altaffilmark{1} and B. Wang\altaffilmark{2,3}}
\email{zwliu@ynao.ac.cn} 
\altaffiltext{1}{Argelander-Institut f\"ur Astronomie, Universit\"at Bonn, Auf dem H\"ugel 71, D-53121 Bonn, Germany}
\altaffiltext{2}{Key Laboratory for the Structure and Evolution of Celestial Objects, Chinese Academy of Sciences, Kunming 650011, P.R. China}
\altaffiltext{3}{Yunnan Observatories, Chinese Academy of Sciences, Kunming 650011, P.R. China}

\begin{abstract}
Type Iax supernovae (SNe Iax) are proposed as one new sub-class of SNe Ia since they present observational
properties that are sufficiently distinct from the bulk of SNe Ia. SNe Iax are the most common of all types of 
peculiar SNe by both number and rate, with an estimated rate of occurrence of about $5$--$30\%$ of the 
total SN Ia rate. However, the progenitor systems of SNe Iax are still uncertain. Analyzing pre-explosion 
images at SN Iax positions provides a direct way to place strong constraints on the nature of progenitor 
systems of SNe Iax. In this work, we predict pre-explosion properties of binary companion stars in a variety of
potential progenitor systems by performing detailed binary evolution calculations with the 
one-dimensional stellar evolution code {\tt STARS}. This will be helpful for constraining progenitor 
systems of SNe Iax from their pre-explosion observations. With our binary evolution calculations, it is found that the non-degenerate helium (He) 
companion star to both a massive C/O WD ($>1.1\,\rm{M_{\odot}}$) and a hybrid C/O/Ne WD can provide an explanation for 
the observations of SN~2012Z-S1, but the hybrid WD+He star scenario is more favorable. 



\end{abstract}


\section{INTRODUCTION}
 \label{sec:introduction}

Type Iax supernovae (SNe Iax) were recently proposed as one of the largest classes of peculiar SNe \citep{Fole13}.
Members of this subclass, have several observational similarities to normal SNe Ia \citep{Li03, Bran04, Jha06, 
Phil07, Fole09, Fole10, Fole13}, but also present sufficiently distinct observational properties.  
SNe Iax are significantly fainter than normal SNe Ia \citep{Fole13, Fole14}. They have a wide range of explosion 
energies ($10^{49}$--$10^{51}\,\rm{erg}$), ejecta masses ($0.15$--$0.5\,\rm{M_{\odot}}$), and $^{56}\rm{Ni}$ 
masses ($0.003$--$0.3\,\rm{M_{\odot}}$). The spectra of SNe Iax are characterized by lower expansion 
velocities ($2000$--$8000\,\rm{km\,s^{-1}}$) than those of normal SNe Ia ($15,000\,\rm{km\,s^{-1}}$) at similar epochs.
Moreover, two SNe Iax (SN~2004cs and SN~2007J) were identified with strong He lines in their spectra \citep{Fole09, Fole13}.

SNe Iax are the most common of all types of peculiar SNe by both number and rate, with an estimated rate of occurrence 
of about $5\%$--$30\%$ of the total SN Ia rate \citep{Li11, Fole13}. Most SNe Iax are observed in late-type, 
star-forming galaxies \citep{Fole13, Lyma13, Whit14}.
The observational constraints suggest relatively short delay times for the progenitor systems of SNe Iax ($<500\,\rm{Myr}$, 
see \citealt{Fole14}). However, there is at least one Iax event, SN 2008ge, which was observed 
in an S0 galaxy with no signs of star formation (which suggests a long delay time, see \citealt{Fole10, Fole13}).

It is accepted that SNe Iax are either thermonuclear explosions of accreting white dwarfs (WDs) or the core-collapse 
of massive stars. However, the nature of the progenitor systems which give rise to SNe Iax has not 
been yet elucidated \citep{Fole13, Fole14, Liu13b, Liu15, McCu14, Fish15}. Given the diversity of SNe Iax,
multiple different progenitor channels are likely to contribute to the observed 
population. The most likely progenitor scenarios currently proposed for SNe Iax are summarized 
as follows.

$\textit{1.\,The Chandrasekhar-mass (Ch-mass) scenario}$. A WD accretes 
hydrogen (H)- or helium (He)-rich material from a 
non-degenerate companion until its mass approaches the Ch-mass, $M_{\rm{Ch}}\approx1.4\,\rm{M_{\odot}}$, 
at which point a thermonuclear explosion ensues \citep{Whel73, Nomo82, Nomo84}.  Here, the companion 
star could be a H-rich donor (either a main-sequence [MS] star, a subgiant, or a red giant [RG]) or 
a He-rich donor (He star). It has been suggested that a Ch-mass WD can undergo 
either a deflagration, or a detonation, or a delayed detonation to lead to a SN 
Ia explosion \citep{Arne69, Woos86, Khok89}. 
In this work, we only focus on SNe Iax that were recently
thought to likely be produced from the deflagration explosion Q2
of a WD. Turbulent deflagrations in WDs can cause strong 
mixing of the SN ejecta (e.g. \citealt{Roep05}), and the small amount of kinetic 
energy released from deflagration explosions is in good agreement with the low 
expansion velocities of SNe Iax \citep{Fole13}. 
Furthermore, the (weak) pure deflagration explosion of a Ch-mass WD has recently been proposed as 
a promising scenario for SNe Iax \citep{Krom13, Fink14, Krom15}.
In this specific deflagration explosion scenario, a weak deflagration explosion fails to explode the entire WD, leaving 
behind a polluted WD remnant star \citep{Jord12, Krom13, Fink14}.
Hydrodynamical simulations have shown that the off-centerignited
weak deflagrations of Ch-mass C/O or hybrid C/O/Ne
WDs \citep{Garc97, Chen14, Deni15} 
are able to reproduce the characteristic observational features of SNe Iax quite 
well \citep{Jord12, Krom13, Fink14, Krom15}, including the faint 2008ha-like SNe Iax \citep{Krom15}.
However, it is impossible to determine
the exact ignition condition of a Ch-mass WD in this work because the WD is treated 
as a point mass in our binary calculations (see Section~\ref{sec:code}). We 
therefore assume that all Ch-mass WDs in our binary calculations will lead to (weak) pure 
deflagration explosions for SNe Iax.

In addition, the pulsational delayed detonation explosion of a near Ch-mass WD has been 
proposed to be a possible scenario for the SN Iax SN~2012Z \citep{Stri14} which
has been suggested to come from a progenitor system consisting of a He-star donor and a C/O WD \citep{McCu14}.
However, further and more detailed hydrodynamic and radiative-transfer simulations for this pulsational delayed detonation 
explosion model are still needed.

$\textit{2.\,The sub-Ch-mass scenario}$ (e.g. \citealt{Iben87, Livn90, Woos94, Woos11}). The 
WD accretes He-rich material from a companion star (a non-degenerate He star or a He WD) to 
accumulate a He-layer on its surface.  If an He-shell accumulation reaches a critical value 
($\simeq0.02$--$0.2\,\rm{M_{\odot}}$), a single-detonation (i.e., ``.Ia model,'' see \citealt{Bild07, Shen10}) 
or double-detonation is triggered to cause the thermonuclear explosion of the sub-Ch-mass WD. However, current works show that 
theoretical spectra and light-curves predicted from the simulations with a thick He shell ($0.1$--$0.2\,\rm{M_{\odot}}$) 
do not match the observations of SNe Ia/Iax \citep{Krom10, Woos11, Sim12}. Therefore, we will assume a thin
He shell ($\approx0.05\,\rm{M_{\odot}}$) in our subsequent calculations for the sub-Ch-mass scenario. 
Recent population synthesis calculations for the 
sub-Ch-mass double-detonation scenario suggest that the rates and delay-time distribution 
from this scenario seem to be consistent with the observations of SNe Iax \citep{Ruit11, Wang13}. However, 
current explosion models for the double-detonation scenario \citep{Bild07, Fink10, Krom10, Sim10, Shen10, Woos11} 
show that this kind of explosion struggles to reproduce the characteristic features of SNe Iax, for example, 
the low ejecta velocity of SNe Iax, low $\rm{^{56}Ni}$ mass, and strong mixing of their explosion ejecta.
Also, the sub-Ch-mass explosions are not going to undergo a weak pure deflagration as in the case for 
Ch-mass WDs. Nevertheless, further investigations are still required and numerous complications remain to be 
solved in such a model \citep{Pier14, Shen14, Piro15}. For instance, \citet{Piro15} 
suggests that including turbulent mixing in He-accreting WDs can lead to more than $50\%$ C/O in the accreted layer at the time of
ignition, which probably causes the nucleosynthetic products to be different, or similar.
It is unclear if any extended explosion model within this scenario can successfully reproduce most 
observational features of the least luminous SNe Iax. We therefore also include the sub-Ch-mass explosion as a potential 
scenario for producing SNe Iax in our presented calculations.

$\textit{3.\,The massive-star core-collapse scenario}$ (e.g. \citealt{Umed05, Vale08, Mori10}). A core-collapse 
SN is triggered by the gravitational collapse of the Fe core of a massive star (with an initial MS mass above
about $10\,\rm{M_{\odot}}$. that had its H and a significant amount of He stripped from its outer layers.
In particular, the fallback core-collapse explosions of massive stars have been proposed to explain the peculiar SN Iax 
SN~2008ha because these specific SN explosions could produce a low explosion energy and $\rm{^{56}Ni}$ mass 
to account for the properties of SN~2008ha \citep{Mori10}. However, this core-collapse scenario seems difficult 
to explain the full diversity of SNe Iax. Also, a TP-AGB-like source at the position of SN~2008ha 
has been detected by a recent $\textit{Hubble Space Telescope (HST)}$ analysis \citep{Fole14}. This 
source has been further suggested to be a candidate of
either the bound remnant of the WD, or the companion star of the WD, or the companion of the massive-star 
progenitor (although it is very unlikely, see \citealt{Fole14}).  If future observations confirm that this source is 
indeed a bound remnant of the WD, a massive star as the progenitor of SN~2008ha would be ruled out.
Also, pre-explosion imaging for another SN Iax, SN~2008ge, has ruled out particularly massive stars as potential
progenitors \citep{Fole10, Fole15}.

Analyzing pre-explosion images at the SN position provides a direct way to constrain SN progenitor 
systems (e.g., see \citealt{Fole14, Fole15, McCu14}). To date, no progenitors of normal SNe Ia have 
yet been directly observed, even for detections of relatively nearby
SNe Ia, SN~2011fe and SN~2014J. However, the probable progenitor system of a SN Iax
SN~2012Z (i.e., SN~2012Z-S1) has been recently discovered from pre-explosion $HST$ 
images \citep{McCu14}. It is further suggested that SN~2012Z had a progenitor system consisting of 
a He star donor and a C/O WD \citep{McCu14}.

In the accreting-WD explosion scenario, the WD can only be observed directly in our
own Milky Way and several very nearby galaxies because the WD would be faint. Consequently, the companion 
stars generally play a major role in determining the pre-explosion signatures of progenitor systems. 
It may be possible to determine the nature of the companion star of SNe Iax, and thus 
their progenitors, by analyzing pre-explosion images at the SN positions. In this work, we predict 
pre-explosion signatures of companion stars for different progenitor systems that have been recently 
proposed for SNe Iax. This will be very helpful for analyzing future pre-explosion observations of SNe Iax. 
In Section~\ref{sec:code}, we describe the numerical 
method we used in this work. The 
pre-explosion signatures of companion stars are presented in Section~\ref{sec:results}. 
A comparison between our results and the observation of SN~2012Z-S1 is presented in Section~\ref{sec:12z}.
The conclusions are summarized in Section~\ref{sec:summary}.

\section{Numerical method and models}
\label{sec:code}

We use the Cambridge stellar evolution code {\tt STARS}, which was originally written 
by \citet{Eggl71, Eggl72, Eggl73} and then updated several times \citep{Han00}. The detailed 
description of the version we used can be found in \citet{Han00} and \citet{Han04}. Generally, Roche-lobe 
overflow (RLOF) is treated by the prescription of \citet{Han00}. All our stars have 
metallicity $\rm{Z=0.02}$ and start in circular orbits. For H-rich stars (MS, subgiant, 
and RG),  we set the ratio of the typical mixing length ($l$) to the local pressure scale 
height ($H_{P}$), $\alpha=l/H_{P}=2.0$, and the convective overshooting parameter, $\delta_{\mathrm{ov}}=0.12$, 
which roughly corresponds to an overshooting length of $\approx0.25\,H_{P}$. 
The He-rich star is evolved without enhanced mixing, i.e., the convective overshooting 
parameter, $\delta_{\mathrm{ov}}=0$.


In our binary evolution calculations, the accreting WDs (i.e., C/O WDs or hybrid C/O/Ne WDs) are 
treated as a point mass, only detailed structures of 
the mass donor stars are solved in the code. We trace detailed mass-transfer of the binary systems 
until the WDs increase their mass to the critical limit (assuming SN Iax explosions occur). For different 
companion star models, the mass-growth rate of the WD, $\dot{M}_{\rm{WD}}$, is defined as follows.

\begin{figure*}
  \begin{center}
    {\includegraphics[width=0.45\textwidth, angle=360]{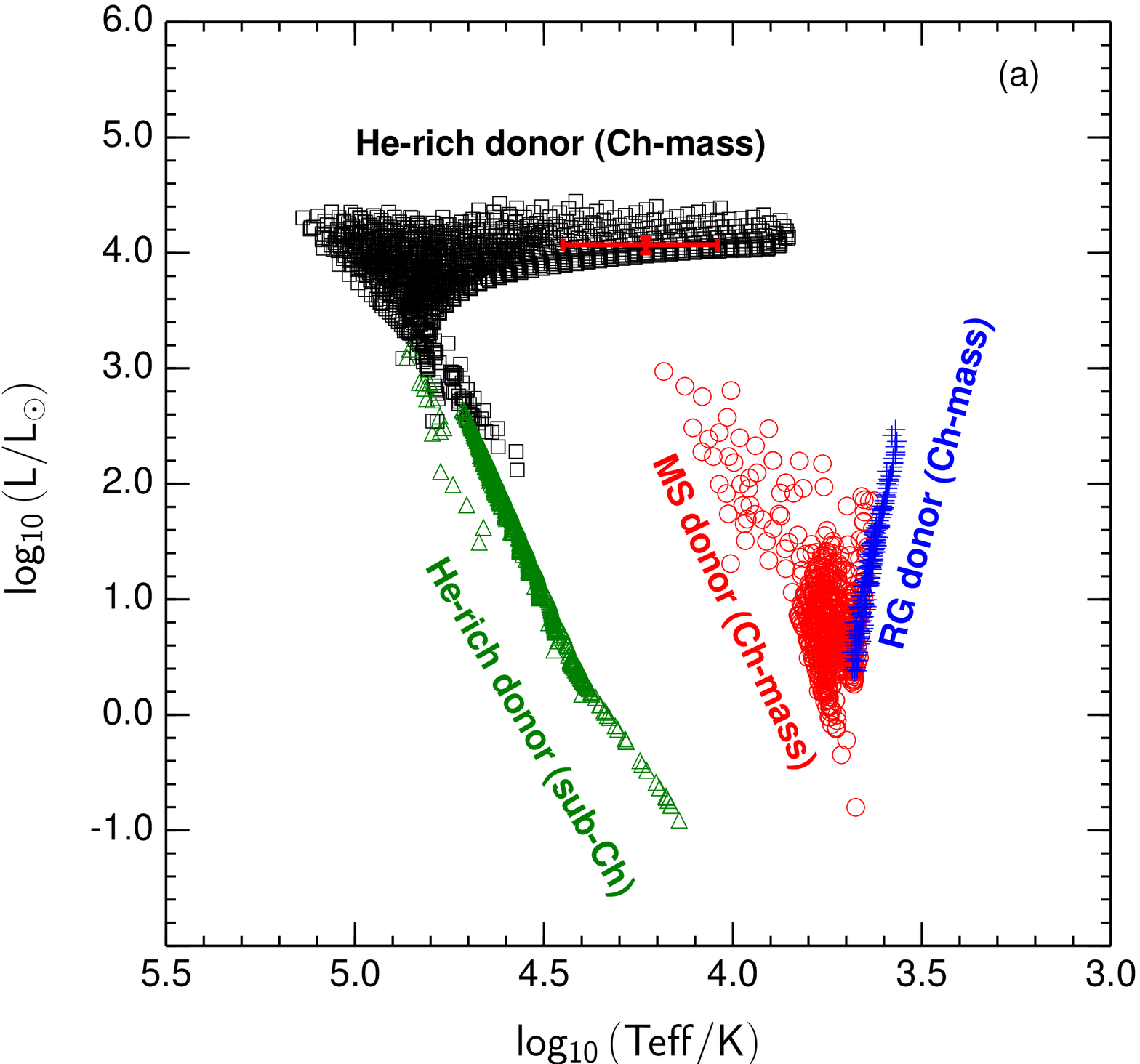}}
    {\includegraphics[width=0.45\textwidth, angle=360]{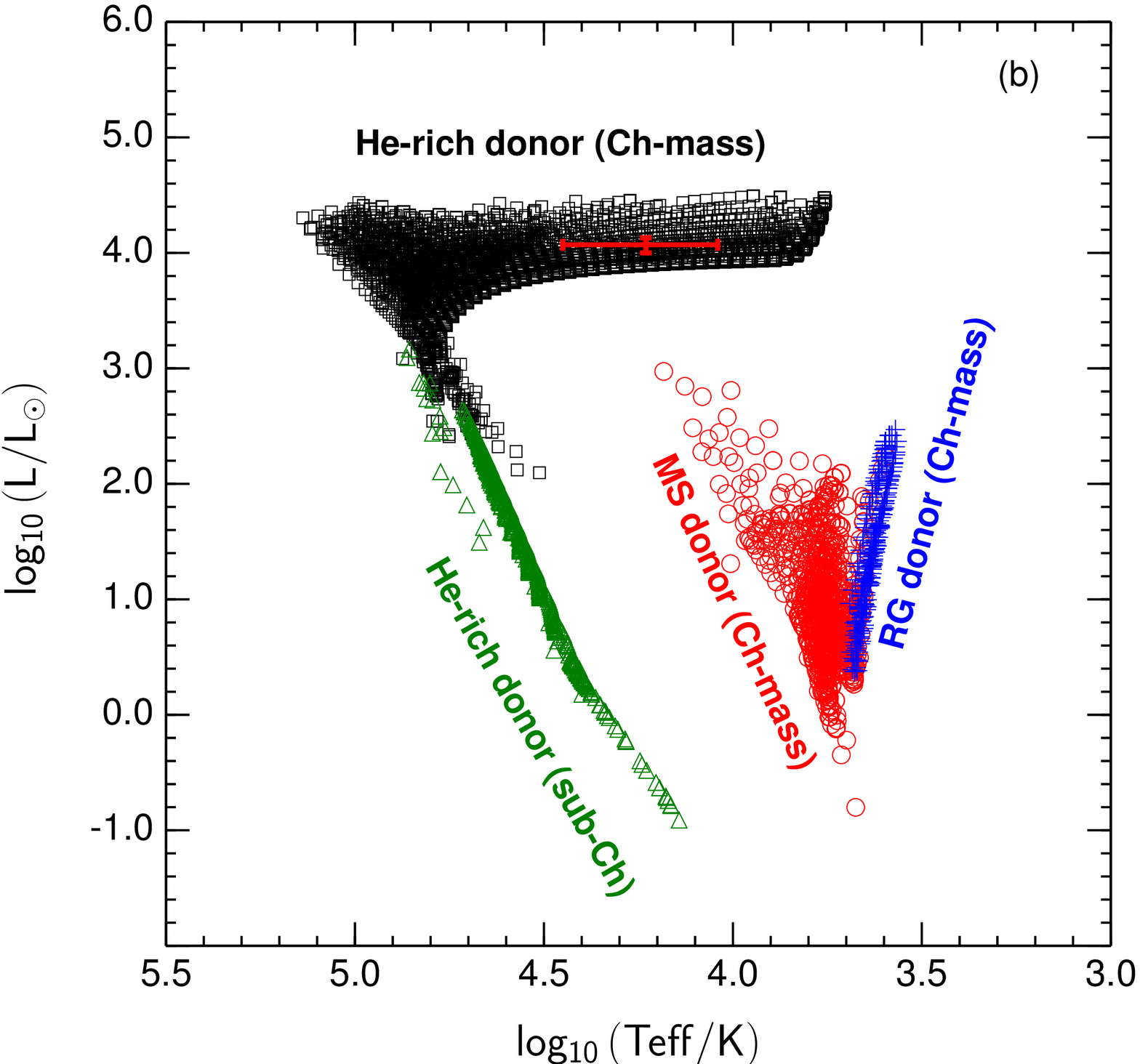}}
    {\includegraphics[width=0.45\textwidth, angle=360]{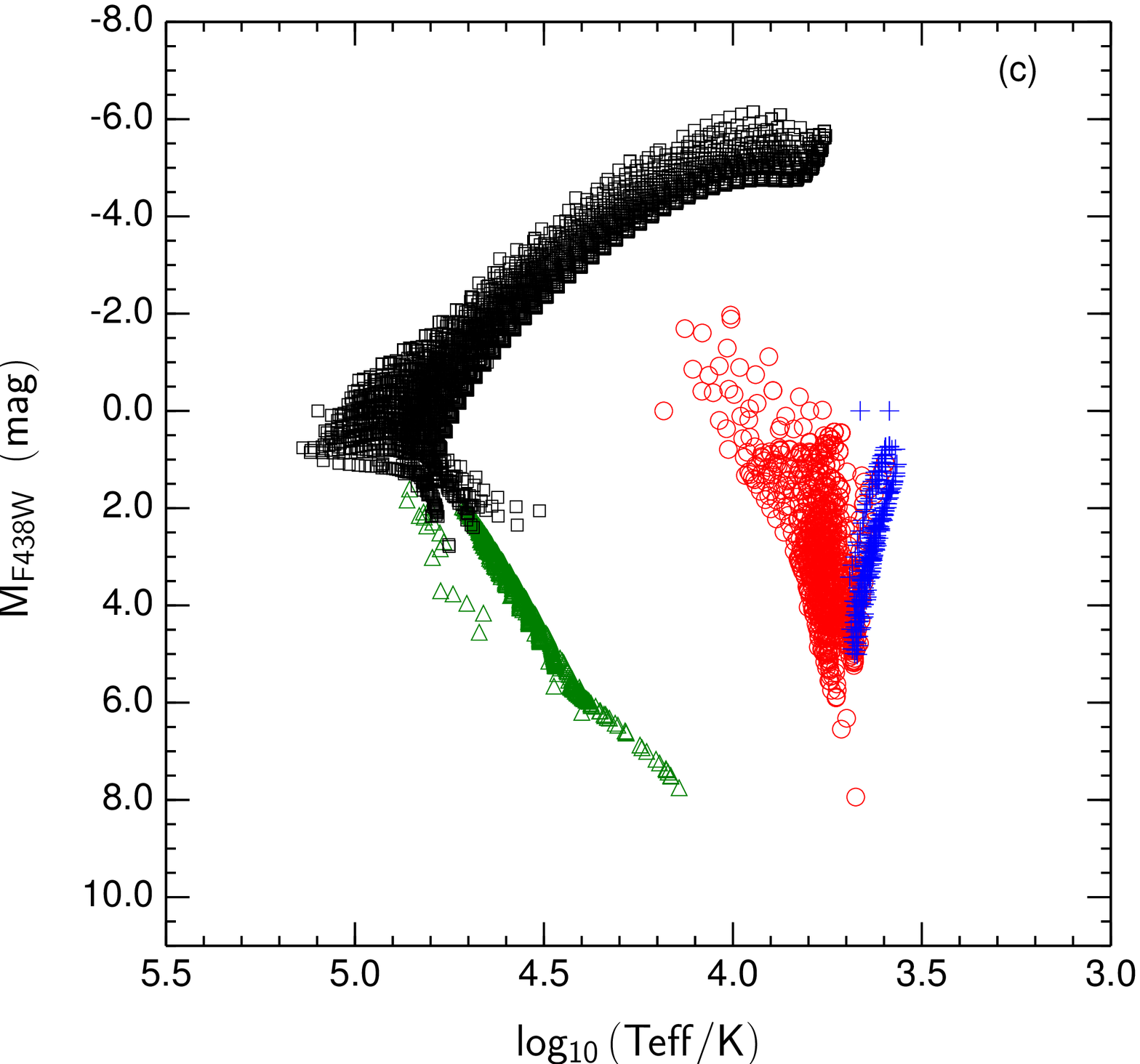}}
    {\includegraphics[width=0.45\textwidth, angle=360]{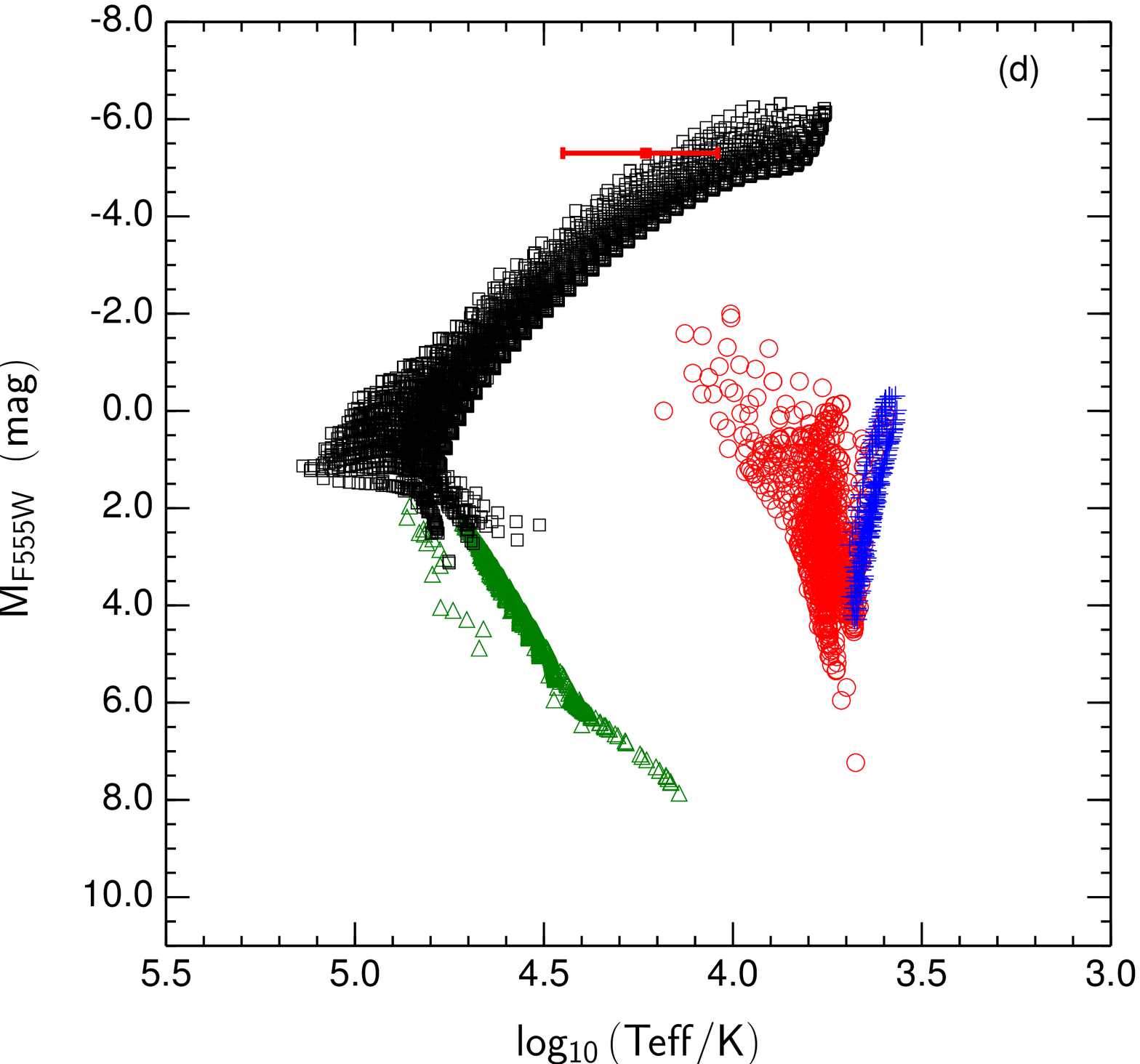}}
    {\includegraphics[width=0.45\textwidth, angle=360]{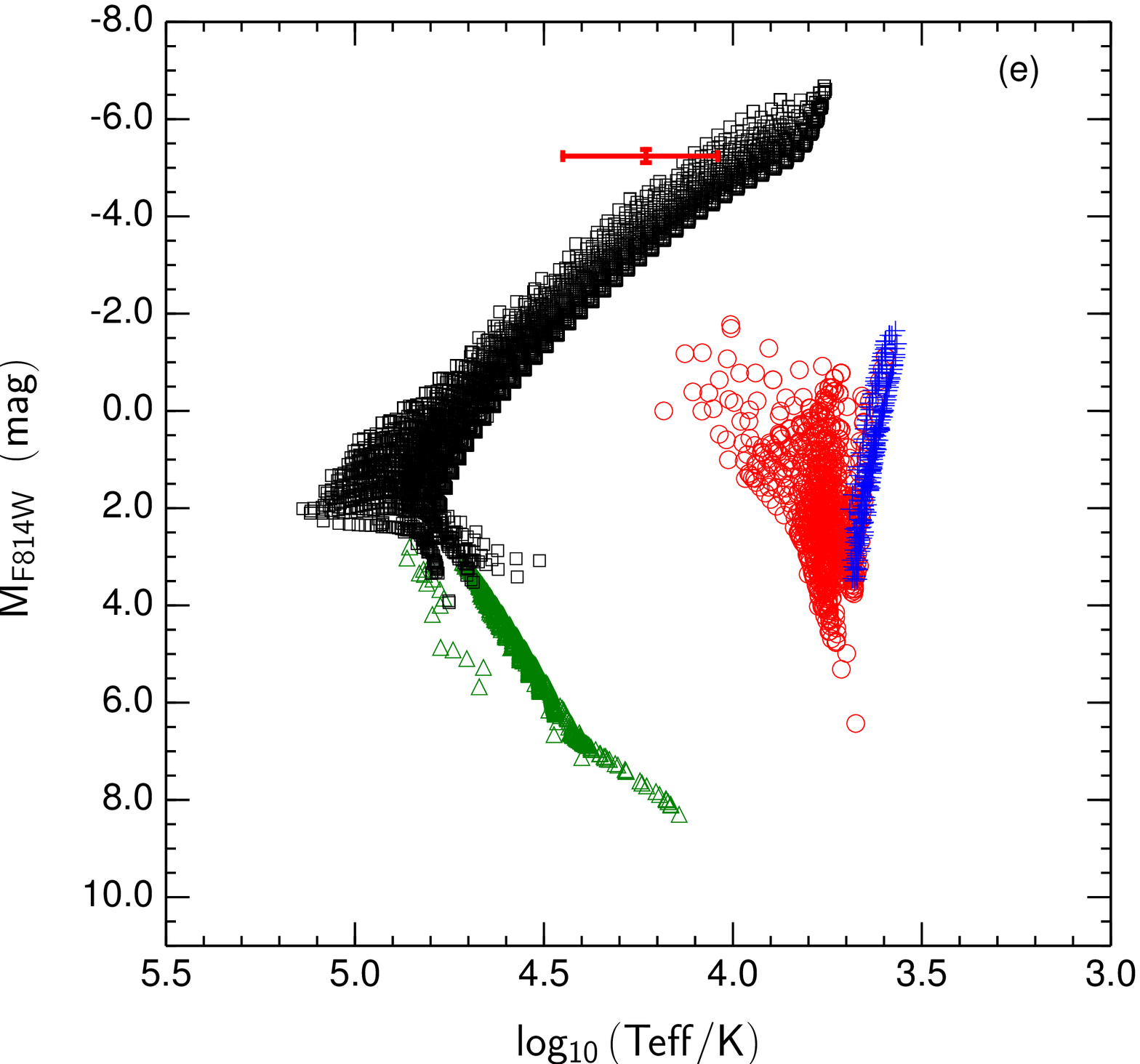}}
    {\includegraphics[width=0.45\textwidth, angle=360]{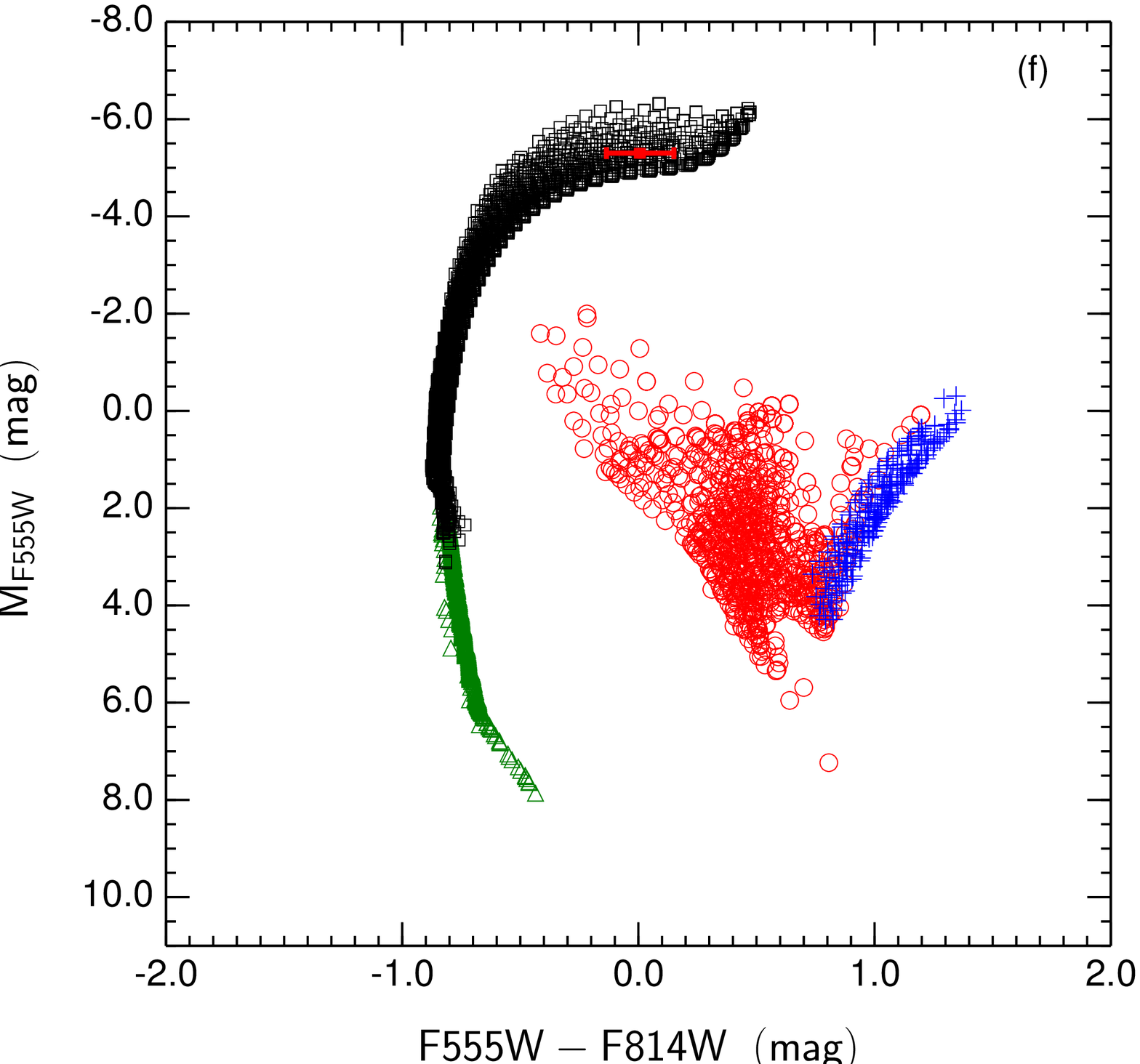}}
  \caption{Panel (a): H-R diagram of pre-explosion companions of SNe Iax in different progenitor scenarios. In this panel, initial 
           masses of WDs in the Ch-mass scenario up to a maximum value of $\rm{1.2\,M_{\odot}}$. Panel (b): similar to panel (a), but $\rm{1.3\,M_{\odot}}$ 
           initial mass WDs are also included for the H/He-rich accreting Ch-mass scenario. Panel (c),(d) and (e): similar to panel(b), but for different 
           HST/WFC3 band magnitude vs. effective temperature. Panel (f): similar to panel (b), but for colour-magnitude diagram ($\rm{M_{F555W}}$-$\rm{M_{F814W}, M_{F555W}}$).
           The red error bar shows the $\textit{HST}$ observation of the SN~2012Z-S1 of \citet{McCu14}. The red circle, blue cross, and black square symbols
           corresponding to the MS, RG, and He star donor Ch-mass channel. The green triangle symbols represent the sub-Ch-mass channel.
            }
\label{Fig:1}
  \end{center}
\end{figure*}

$\textit{For the H-rich donor star channel}$, the mass growth rate of the WD, 
$\dot{M}_{\rm{WD}}=\eta_{\rm{He}} \dot{M}_{\rm{He}} = \eta_{\rm{He}}\eta_{\rm{H}}|\dot{M}_{\rm{tr}}|$,  
where  $\dot{M}_{\rm{He}}$ is the mass-growth rate of the He layer under the H-shell burning, and $|\dot{M}_{\rm{tr}}|$ is the mass-transfer 
rate from the non-degenerate H-rich companion star. Also, ${\eta}_{\rm{He}}$  is the mass-accumulation efficiency for 
He-shell flashes, which is shown below in detail (see also \citealt{Kato04}), and $\eta_{\rm{H}}$ 
is the mass-accumulation efficiency for H burning, which is defined as follows (see also \citealt{Liu15}).\footnote{
Here, the effect of the accretion disk around a WD is not considered because whether a disk instability occurs or not 
during the WD-accreting process is unclear (see \citealt{Hach10, Kato12}).}

 \begin{equation}
    \label{eq:a1}
\eta_{\rm{H}} = \left\{ \begin{array}{ll}
\dot{M}_{\rm{cr,H}}/|\dot{M}_{\rm{tr}}|,  & |\dot{M}_{\rm{tr}}| > \dot{M}_{\rm{cr,H}},  \\
1, & \dot{M}_{\rm{cr,H}} \geqslant |\dot{M}_{\rm{tr}}| \geqslant\frac{1}{8}\dot{M}_{\rm{cr,H}},\\
0, & |\dot{M}_{\rm{tr}}| < \frac{1}{8}\dot{M}_{\rm{cr,H}},
\end{array} \right.
  \end{equation}
where $\dot{M}_{\rm{cr,H}}=5.0\times10^{-7}(1.7/X-1)(M_{\rm{WD}}/\rm{M_{\odot}}-0.4)\,\rm{M_{\odot}yr^{-1}}$ is 
the critical accretion rate for stable H burning, $X$ is the H mass fraction, $M_{\rm{WD}}$ the mass of the accreting WD. 
Here, the optically thick wind \citep{Hach96} 
is assumed to blow off all unprocessed material if $|\dot{M}_{\rm{tr}}|$ is greater than $\dot{M}_{\rm{cr,H}}$, and the lost 
material is assumed to take away the specific orbital angular momentum of the accreting WD. The effect of magnetic braking 
is also included by adopting the description of angular-momentum loss from \citet{Sill00}. If 
the C/O WD increase its mass to near the Ch-mass limit ($\approx1.4\,\rm{M_{\odot}}$), we assume 
the SN Iax explosion ensues. 

Very recently, considering the uncertainties of the C-burning rate, \citet{Chen14} have suggested that 
hybrid C/O/Ne WDs as massive as $1.3\,\rm{M_{\odot}}$ can be formed if the convective undershooting of the C-burning
layer is large enough, the carbon fraction below the flame is largely reduced. These hybrid  C/O/Ne WDs in close 
binary systems can eventually accrete 
H- or He-rich material from their companion stars 
to reach the Ch-mass limit and explode as faint SNe (see \citealt{Garc97, Chen14, Deni15}). 
In this work, this hybrid WD scenario has also been included in our detailed 
binary evolution calculations for the Ch-mass scenario. Because all WDs are treated as a point mass in our calculations, the C/O WD and 
the hybrid WD channel are distinguished just based on their initial WD mass (see Section~\ref{sec:12z}). The 
initial WD masses of $\rm{M_{WD}^{i}=0.65, 0.7, 0.8, 0.9, 1.0, 1.1, 1.2}$ and $\rm{1.3\,M_{\odot}}$ 
are considered for the H-accreting 
Ch-mass scenario (see also \citealt{Liu15}). With a low initial WD mass of $0.65\,\rm{M_{\odot}}$, only two systems (out of about 
120 binary systems that are calculated)
can lead to SN Iax explosions, $0.65\,\rm{M_{\odot}}$ is thus set to be the minimum WD mass for producing SNe Iax 
from this channel. A WD mass of $1.3\,\rm{M_{\odot}}$ is the maximum initial mass we assumed for a hybrid C/O/Ne 
WD (see \citealt{Chen14, Deni15}).

$\textit{For the He-rich donor star channel}$, the mass growth rate of the WD, $\dot{M}_{\rm{WD}}=\eta_{\rm{He}} |\dot{M}_{\rm{tr}}'|$ (see 
also \citealt{Wang10, Liu15}), where $|\dot{M}_{\rm{tr}}'|$ is the mass-transfer rate from the He-rich companion star, the $\eta_{\rm{He}}$ is 
the mass-accumulation efficiency for He burning, which is set to be 
 
 \begin{equation}
    \label{eq:a3}
\eta_{\rm{He}} = \left\{ \begin{array}{ll}
\dot{M}_{\rm{cr,He}}/|\dot{M}_{\rm{tr}}'|,  & |\dot{M}_{\rm{tr}}'| > \dot{M}_{\rm{cr,He}},  \\
1, & \dot{M}_{\rm{cr,H}} \geqslant |\dot{M}_{\rm{tr}}'| \geqslant\dot{M}_{\rm{st}},\\
\eta_{\rm{He}}', & \dot{M}_{\rm{st}} \geqslant |\dot{M}_{\rm{tr}}'| \geqslant\dot{M}_{\rm{low}},\\
1\,(\rm{no\ burning}), & |\dot{M}_{\rm{tr}}'| < \dot{M}_{\rm{low}},
\end{array} \right.
  \end{equation}
where $\dot{M}_{\rm{cr,He}} = 7.2\times10^{-6}(M_{\rm{WD}}/\rm{M_{\odot}}-0.6)\,\rm{M_{\odot}yr^{-1}}$ is 
the critical accretion rate for stable He burning; $\dot{M}_{\rm{st}}$ is the minimum accretion
rate for stable He-shell burning \citep{Kato04, Pier14}; $\dot{M}_{\rm{low}}=4\times10^{-8}\,\rm{M_{\odot}yr^{-1}}$ 
is the minimum accretion rate for weak He-shell flashes \citep{Woos86}; ${\eta}_{\rm{He}}'$ is obtained from 
linearly interpolated from a grid computed by \citet{Kato04}. As above mentioned, if the WD 
increases its mass to the near Ch-mass limit ($\approx1.4\,\rm{M_{\odot}}$), we assume the SN explosion ensues. 
Here, the initial WD masses of $\rm{M_{WD}^{i}=0.865, 0.9, 1.0, 1.1, 1.2}$ and $\rm{1.3\,M_{\odot}}$
are considered for the He-accreting Ch-mass scenario (see also \citealt{Wang10, Liu15}). The lower 
mass limit of $0.865\,\rm{M_{\odot}}$ is corresponding to the minimum WD mass for producing SNe Iax from 
this channel, and a upper-limit of $1.3\,\rm{M_{\odot}}$ is the maximum initial mass we assumed for the 
hybrid C/O/Ne WD (see \citealt{Chen14, Deni15}).

\textit{For low mass-transfer-rate He-accreting case, $|\dot{M}_{\rm{tr}}'|\lesssim 4\times10^{-8}\,\rm{M_{\odot}yr^{-1}}$}, 
a thick layer of helium is believed to grow on the surface of the WD. As a result, once an He-shell accumulation 
reaches a critical value ($0.05\,\rm{M_{\odot}}$ is adopted in this work, see \citealt{Woos11}), we 
assume a single detonation or double detonation is 
triggered, the sub-Ch-mass WD explodes as an SN Iax. When the mass-transfer rate is low, $|\dot{M}_{\rm{tr}}'|\lesssim 1\times10^{-9}\,\rm{M_{\odot}yr^{-1}}$, 
it is suggested that the flash when the He layer ignites is too weak to initiate a carbon 
detonation, which results in only a single He detonation wave propagating outward \citep{Nomo82}.
We do not specially distinguish this low-mass-transfer case in our binary evolution calculations because the condition of He ignition is 
still uncertain \citep{Bild07, Shen10, Woos11, Shen14}. This is different from the setup in \citet{Wang13} who 
did not include the case of  $|\dot{M}_{\rm{tr}}'|\lesssim 1\times10^{-9}\,\rm{M_{\odot}yr^{-1}}$ 
because they only focused on the double-detonation explosions. In the present work, the initial WD masses 
of $\rm{M_{WD}^{i}=0.8, 0.9, 1.0}$ and $\rm{1.1\,M_{\odot}}$ are adopted for the He-accreting sub-Ch-mass scenario. 
Here, $0.8\,\rm{M_{\odot}}$ corresponds to the prediction for the minimum WD mass for carbon 
burning from recent hydrodynamic simulations. The detonation of a C/O WD may not be triggered for 
lower mass \citep{Sim12}. We also expect that the initial C/O WD masses are below $\simeq1.1\,\rm{M_{\odot}}$ 
as more massive WDs are usually formed consisting of O and Ne, i.e., ONe WDs (\citealt{Sies06, Dohe15}, but see \citealt{Ruit13}).

In the He-accreting sub-Ch-mass scenario, the companion star could also be a He WD. However, the long delay 
times of the He-WD donor channel ($\gtrsim800\,\rm{Myr}$, see \citealt{Ruit11}) are inconsistent with the 
observational constraints on the ages of SN Iax progenitors \citep[$<500\,\rm{Myr}$;][]{Lyma13, Fole13, Whit14}, and this channel is generally proposed to produce 
calcium-rich SNe such as SN~2005E \citep{Fole15b}. 
Therefore, the He-WD donor scenario is not addressed in our calculations.

\begin{figure*}
  \begin{center}
    {\includegraphics[width=0.48\textwidth, angle=360]{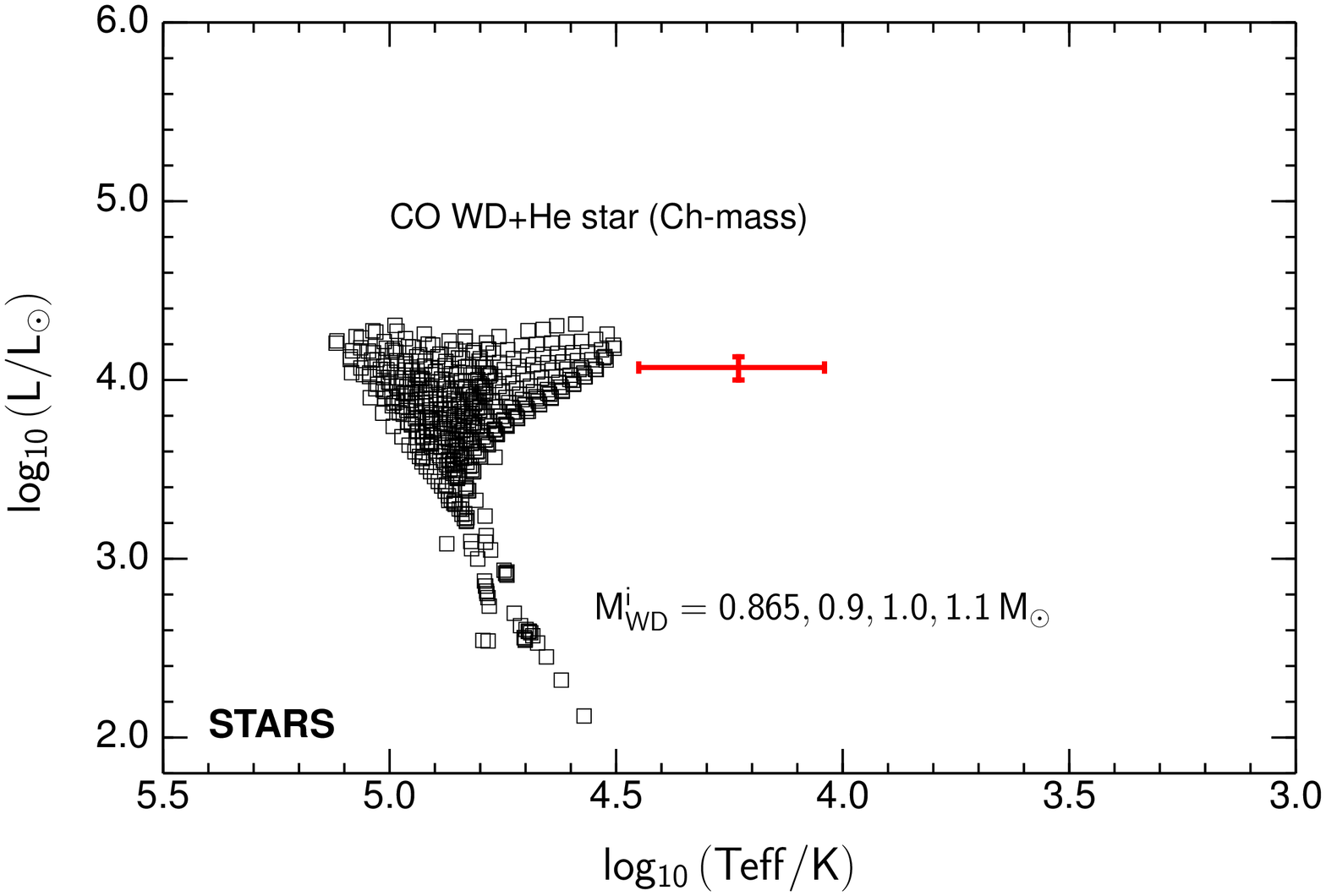}}
    {\includegraphics[width=0.48\textwidth, angle=360]{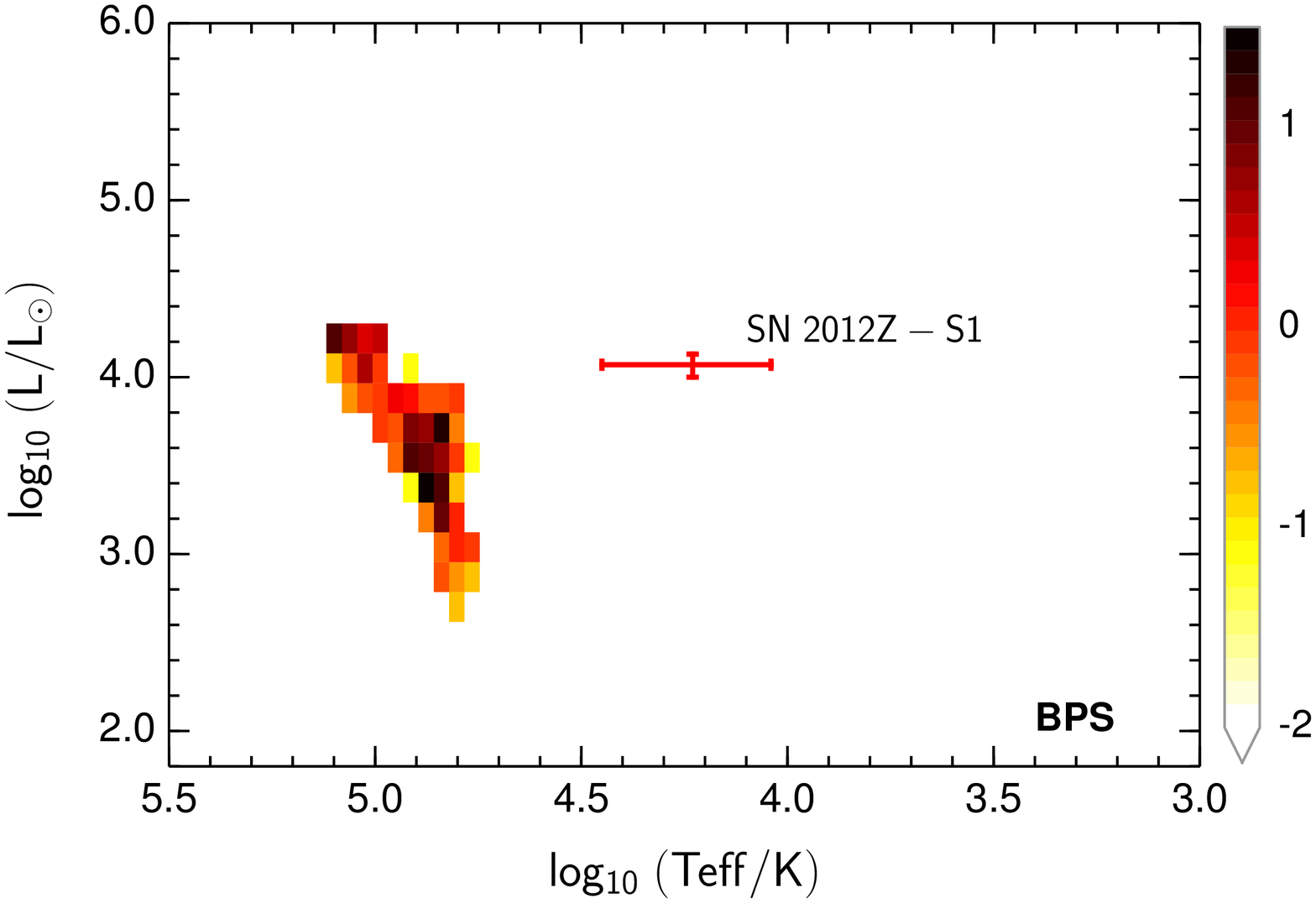}}
    {\includegraphics[width=0.48\textwidth, angle=360]{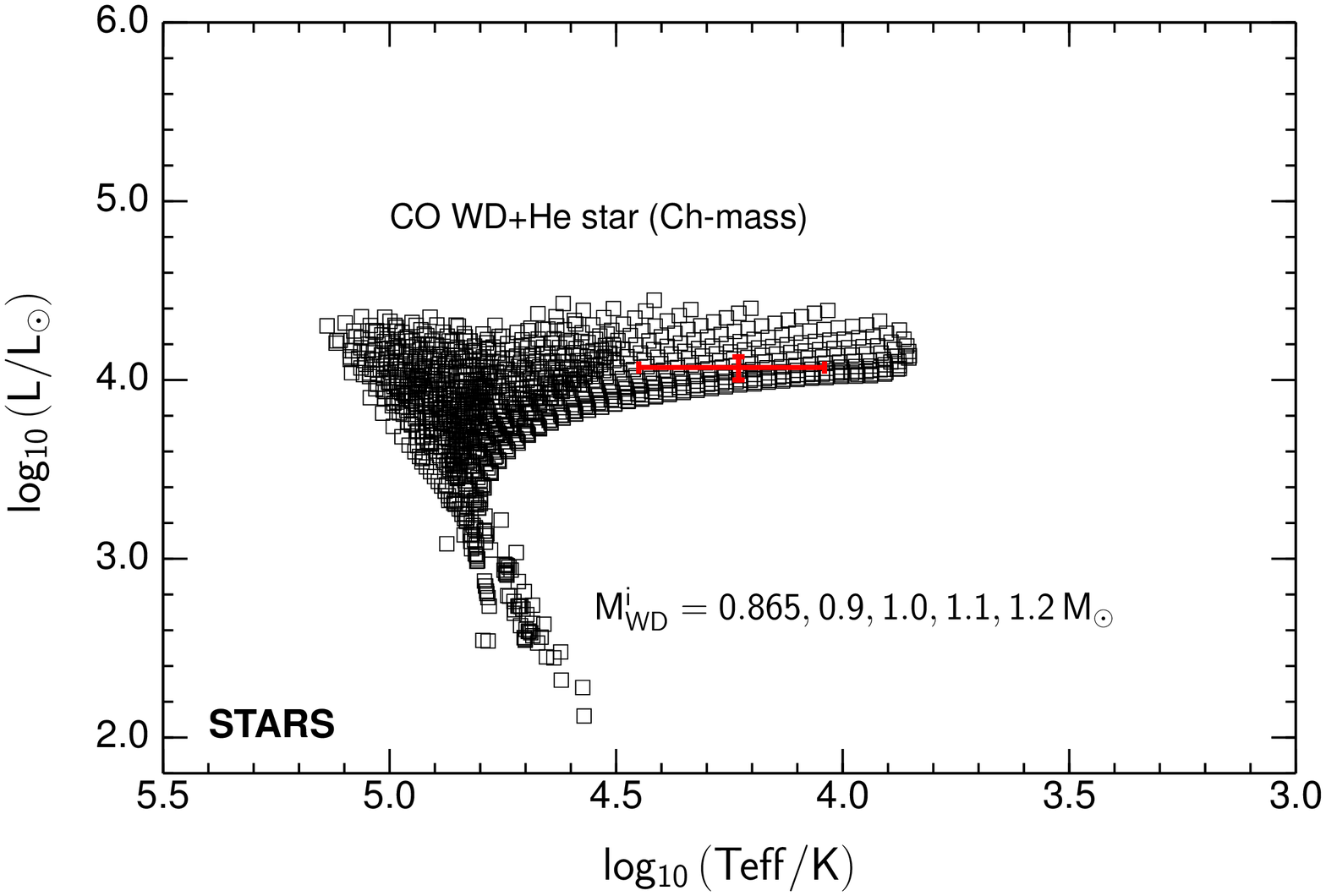}}
    {\includegraphics[width=0.48\textwidth, angle=360]{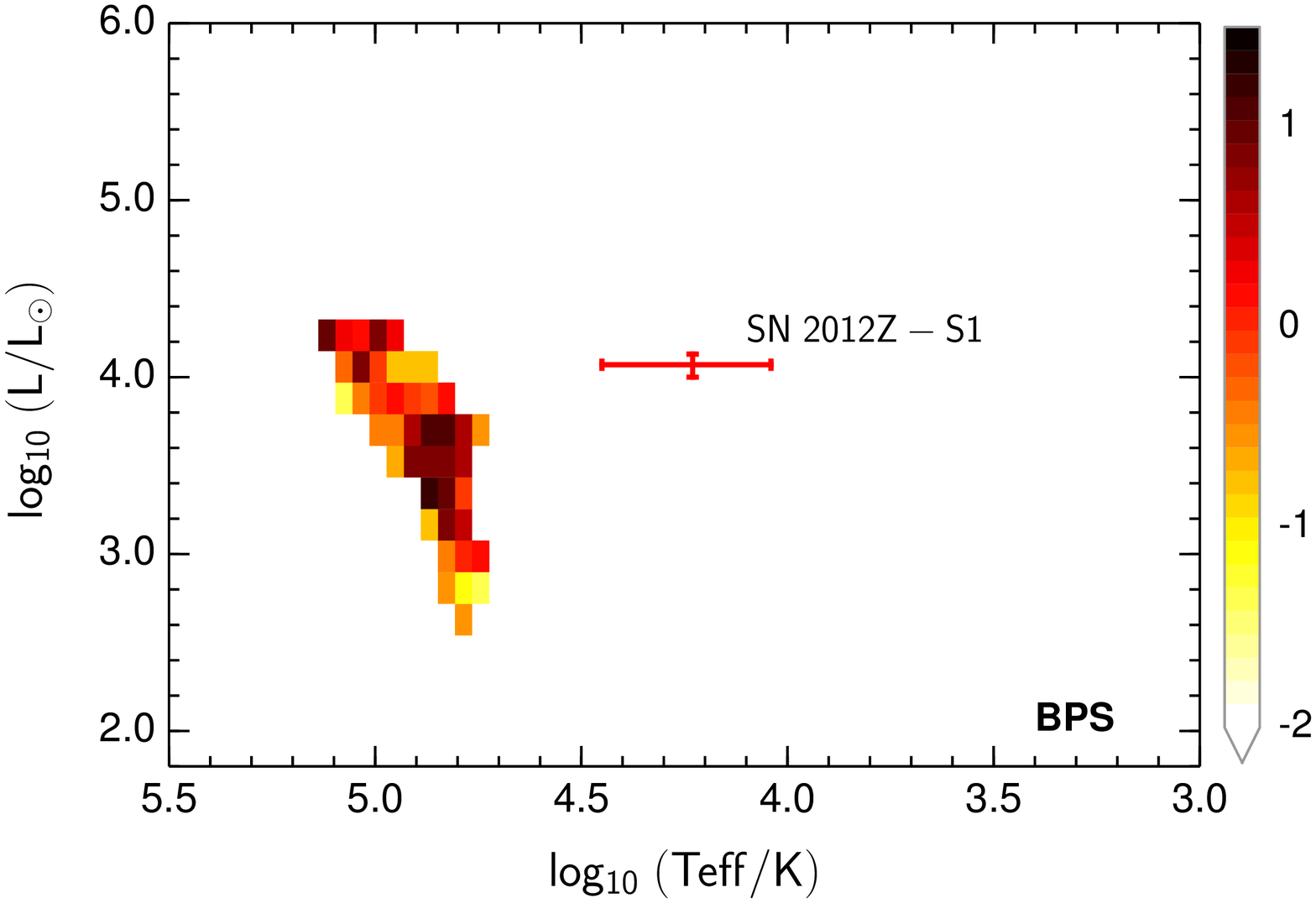}}
    {\includegraphics[width=0.48\textwidth, angle=360]{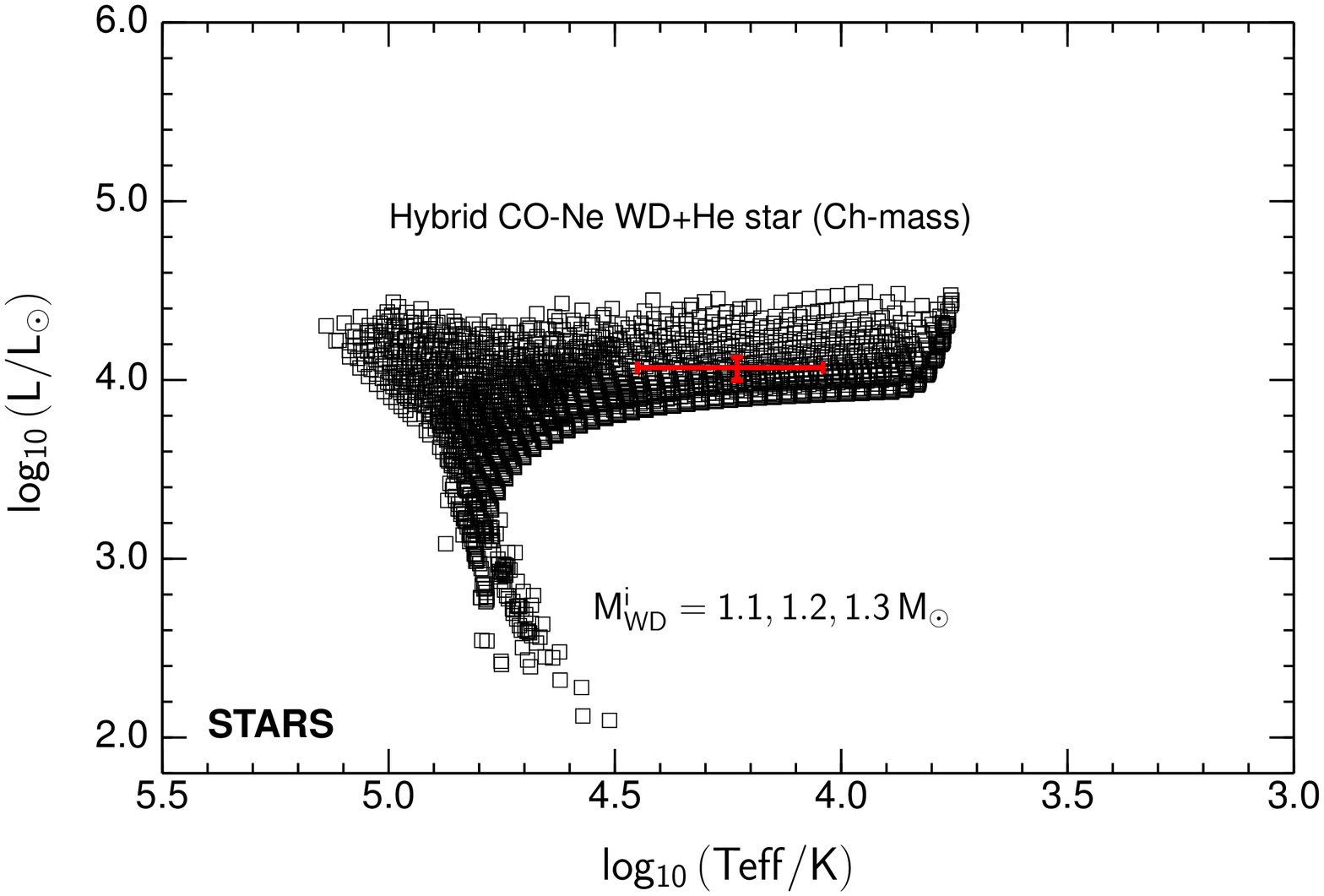}}
    {\includegraphics[width=0.48\textwidth, angle=360]{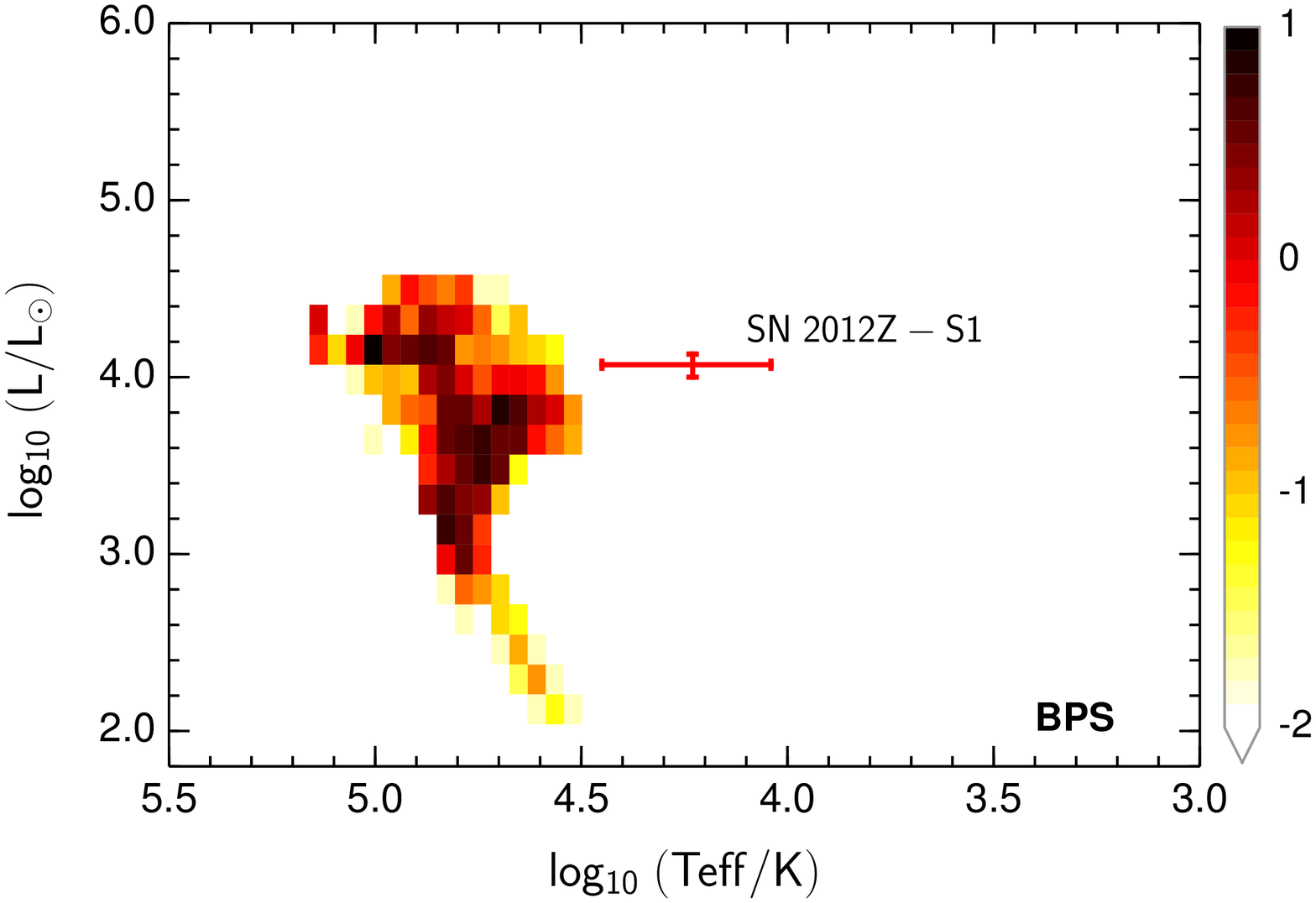}}
  \caption{Left column: similar to panel (b) of Fig.\ref{Fig:1}, but only for the He-star donor Ch-mass 
           scenario. Right column: the distributions (in logarithmic scale) of companion stars in 
           the the plane of ($\rm{log_{10}}$\,$T_{\rm{eff}}$, $\rm{log_{10}}$\,$L$), 
           which are obtained from BPS calculations assuming a constant 
           star formation rate of $3.5\,\rm{M_{\odot}\,\rm{yr^{-1}}}$. Here, the C/O WD+He star (top + middle row) and hybrid C/O/Ne WD+He star (bottom row) 
           Ch-mass channel are considered, respectively. To better compare with the observation
           of the SN~2012Z-S1 (the error in red, see \citealt{McCu14}), we show the results from the C/O WD+He star channel by 
           including (middle row) or excluding (top row) the calculations with an 
           initial WD mass of $1.2\,\rm{M_{\odot}}$. }

\label{Fig:2}
  \end{center}
\end{figure*}

\section{Pre-explosion Companion Stars}
\label{sec:results}
 
Using the method described in Section~\ref{sec:code}, we performed detailed calculations 
for a set of binary systems with various initial WD masses ($M_{\rm{WD}}^{i}$), companion masses ($M_{2}^{i}$) and 
orbital periods ($P^{i}$) in different progenitor scenarios. 
Consequently, the range of companion stars of progenitor properties at the moment of SN Iax explosions, e.g., 
the bolometric luminosity ($L$), effective temperature ($T_{\rm{eff}}$), photospheric radius ($R$), 
orbital velocity ($V_{\rm{orb}}$), and surface gravity ($g$) are directly determined in our binary 
calculations with the {\tt STARS} code. In addition, some properties of binary progenitor systems
when the SN explosion occurs can also be calculated, e.g., the orbital periods of binary systems,
the amount of mass lost during the optically thick stellar wind phase, mass-transfer rate at
the moment of SN explosion, and even equatorial rotational velocities of companion stars (if 
we assume that the companion stars co-rotate with their orbits). All these pre-explosion signatures 
of companion stars and binary progenitors can be compared with pre-explosion observations 
of SNe Iax to help understand the nature of the SN Iax progenitor (e.g., \citealt{McCu14}).

Furthermore, to facilitate direct comparison with optical observations, the bolometric luminosity 
is converted to broad band magnitudes by using the same method as \citet{Pan13} under the
assumption that the companion photosphere emits a blackbody radiation spectrum: 
 
 \begin{equation}
    \label{eq:}
m_{S_{\lambda}} = -2.5\,\rm{log_{10}}\,\left [\frac{\int S_{\lambda} (\pi B_{\lambda})d\lambda}{\int (\textit{f}^{\,0}_{\nu}\,c/\lambda^{2})S_{\lambda}d\lambda}\ \left(\frac{R}{d}  \right)^{2} \right ]  
  \end{equation}

where $S_{\lambda}$ is the sensitivity function of a given filter at wavelength $\lambda$, $B_{\lambda}$ is 
the Planck function, $d$ is the distance of the star, and $f^{0}_{\nu}=3.631\times10^{-20}\,\rm{erg\,cm^{-2}\,s^{-1}\,Hz^{-1}}$ is 
the zero-point value in the AB magnitude system. Here, the effective temperature ($T_{\rm{eff}}$) and 
photospheric radius ($R$) of a pre-explosion companion star obtained from our binary evolution calculations 
are needed. Specifically for this work, the different filters of the HST/WFC3 system and their 
corresponding sensitivity functions are considered in this study. The absolute magnitudes are
calculated in the $AB$ magnitude system.

Figure~\ref{Fig:1} illustrates the Hertzsprung-Russell (H-R) diagram in different representations using 
the HST wavebands.  As it is shown, companion stars in the H-rich donor and He-rich donor
channel are independently located in the H-R diagram at the moment of SN explosion. We can also 
clearly distinguish the location of companion stars between the He-rich donor Ch-mass channel (i.e., C/O WD+He star and hybrid WD+He star channel) and 
the He-rich donor sub-Ch-mass channel. This indicates that we can probably rule out (or confirm) the proposed progenitor
scenarios for SNe Iax with pre-explosion observations at the SN Iax position. But unfortunately, we cannot 
fully distinguish the hybrid C/O/Ne WDs' companions and C/O WDs' companions in the Ch-mass explosion scenarios because
the accreting WDs are treated as a point in our detailed binary evolution calculations (but see the discussions 
in the Section~\ref{sec:12z}).


\section{Comparison with SN~2012Z}
\label{sec:12z}

Recently, a luminous blue source (SN~2012Z-S1) was detected in pre-explosion HST images that
was coincident with the location of SN~2012Z \citet{McCu14}, and the non-degenerate He companion star to 
a C/O WD (originally proposed for SNe Iax by \citealt{Fole13}) 
was suggested to be the likely explanation of the observations \citep{McCu14}. However, they also suggested that 
the possibility of a massive star and an accretion disk around the exploding WD cannot still be excluded (see Fig.~2 of \citealt{McCu14}). 
Here, we compare our model results with the observations of SN~2012Z-S1.

\subsection{Comparison with Detailed Binary Calculations}

A comparison between the results of our binary evolution calculations for different 
potential progenitor channels and the observations of SN~2012Z-S1 are displayed in Fig.~\ref{Fig:1}. 
As shown in this figure, the predictions from the Ch-mass scenario with a He star donor are consistent
with the observations, which agrees with the previous suggestion for the progenitor system of 
SN~2012Z (see \citet{McCu14}).

However, as previously mentioned, it is impossible to distinguish the C/O WDs and the hybrid WDs
in our binary calculations because the accreting WD is set up as a point mass. To further discuss 
the differences of companion stars in the C/O WD+He star and hybrid WD+He star channel, we simply 
assume the range of initial WD mass for the C/O WD+He star channel of $0.865$--$1.2\,\rm{M_{\odot}}$, the range of 
initial WD mass for the hybrid WD+He star channel of $1.1$--$1.3\,\rm{M_{\odot}}$ (left column of Fig.~\ref{Fig:2}, see
also \citealt{Chen14, Deni15}).   
In Section~\ref{sec:bps}, our binary population synthesis (BPS) calculations for the hybrid WD+He star channel will
obtain that most binary systems in this channel have an initial WD mass larger than $1.1\,\rm{M_{\odot}}$, the binary 
systems with a lower initial WD mass are quite rare. In addition, only few C/O WDs are formed with a mass 
exceeding $1.1\,\rm{M_{\odot}}$. This implies our assumptions on the initial masses of C/O WDs and hybrid WDs 
are appropriate.

The left column in Fig.~\ref{Fig:2} shows the locations of companion stars at the moment of SN explosions for 
the C/O WD+He star and hybrid WD+He star Ch-mass scenario, respectively. It is shown that the observation of 
SN~2012Z-S1 is only consistent with the predicted companion locations from the hybrid WD+He star scenario and 
the C/O WD+He star scenario with an initial C/O WD mass of $1.2\,\rm{M_{\odot}}$. Taking the problem of the origin
of very massive C/O WDs into account, our detailed binary evolution calculations seem to disfavor that 
SN~2012Z-S1 is a non-degenerate companion star to a C/O WD, it is more likely to be a He star
with a hybrid C/O/Ne WD. Because the hybrid WDs have much lower C to O abundance 
ratios at the moment of the explosive C ignition than their pure C/O counterparts \citep{Deni15}, which probably
lead to different observational characteristics from those of the Ch-mass C/O WDs after the SN explosions and thus 
being distinguished by spectroscopy observations. Recently, an off-centre 
deflagration in a near Ch-mass hybrid C/O/Ne WD has been simulated by \citet{Krom15}. This showed that deflagrations 
in near Ch-mass hybrid C/O/Ne WDs can explain the faint SN Iax SN~2008ha \citep{Krom15}. However, only a simple
near Ch-mass hybrid C/O/Ne WD model was used, more future works with considering different initial conditions
are needed to explore the deflagration explosions of hybrid WDs.


\subsection{Comparison with BPS Calculations}
\label{sec:bps}

Furthermore, to obtain the distributions of properties of companion stars of 
C/O WD (or hybrid C/O/Ne WD) + He-star Ch-mass scenario at
the moment of SN explosion, we have performed a detailed Monte
Carlo simulation with a rapid population synthesis code \citep{Hurl00, Hurl02}. The code
follows the evolution of binaries with their properties being recorded 
at every step. If a binary system evolves to a C/O WD (or hybrid C/O/Ne WD) + He-star 
system, and if the system, at the beginning of the RLOF phase,
is located in the SN Ia production regions in the plane of
($\rm{log_{10}}$\,$P^{i}$, $M_{2}^{i}$) for its $M_{\rm{WD}}^{i}$ , where $P^{i}$, $M_{2}^{i}$  and $M_{\rm{WD}}^{i}$ are, respectively,
the orbital period, the secondary mass, and the WD's mass of
the C/O WD (or hybrid C/O/Ne WD) + He-star system at the beginning 
of the RLOF, we assume that an SN Ia is resulted, and the properties
of the WD binary at the moment of SN explosion are obtained
by interpolation in the three-dimensional grid ($M_{\rm{WD}}^{i}$, $M_{2}^{i}$, $\rm{log_{10}}$\,$P^{i}$)
of the close WD binaries obtained in our detailed binary 
evolution calculations. We refer to \citet{Han04} and \citet{Liu15} for 
a detailed description about the BPS method used here.

The initial mass function of \citet{Mill79} is used in this work. We assumed a circular binary 
orbit and set up a constant initial mass ratio distribution (i.e., $n({q}')=$constant, 
see \citealt{Duch13}). We assumed a constant star formation 
rate (SFR) of $3.5\,\rm{M_{\odot}yr^{-1}}$. The standard energy equations of \citet{Webb84} were used to calculate the output of 
the common envelope (CE) phase. Similar to our previous studies (e.g., \citealt{Wang10, Liu15}), we use a single
free parameter $\alpha_{\rm{CE}}\lambda$ to describe the CE ejection process, 
and we set $\alpha_{\rm{CE}}\lambda$=1.5. Here, $\alpha_{\rm{CE}}$ is the CE ejection
efficiency, i.e. the fraction of the released orbital energy used to eject the CE; $\lambda$ is a 
structure parameter that depends on the evolutionary stage of the donor star. 
Specifically for the hybrid WDs+He star channel, another parameter of the uncertainty of the carbon
burning rate (CBR, see \citealt{Chen14}) is introduced and it is set to 0.1 (see also \citealt{Wang14}).

The distributions of companion-star temperatures and luminosities at the moment of SN Iax explosion from our BPS calculations 
for the C/O WD+He star and hybrid WD+He star scenario are shown in Fig.~\ref{Fig:2} (right column).
If we assume the initial maximum C/O WD mass is $1.1\,\rm{M_{\odot}}$, we cannot obtain companion stars
as cool as SN~2012Z-S1 in both our detailed binary calculations and BPS calculations.
In addition, although SN~2012Z-S1 lies in the regions that can potentially be reached by our detailed
binary simulations for the $1.2\,\rm{M_{\odot}}$ C/O WD+He star model (middle row in Fig.~\ref{Fig:2}) and for the 
hybrid WD+He star model (bottom row in Fig.~\ref{Fig:2}), SN~2012Z-S1 is still much cooler than 
predictions from our BPS calculations.
In fact, no star has been obtained from our BPS calculations that lies in the region of the location of SN~2012Z-S1.
We have also studied how the different CE efficiencies 
affect the results. Even if various CE efficiencies (i.e., different values 
of $\alpha_{\rm{CE}}\lambda=0.5$--1.5) are adopted, we still cannot obtain the stars as cool as SN~2012Z-S1. 
However, we caution that BPS calculations still have some uncertainties (see below). Also, Fig.~\ref{Fig:2} shows that 
our detailed binary calculations can significantly contribute to the number of systems 
in the vicinity of SN 2012Z.  The discrepancy between the STARS results and the BPS predictions 
is because we do not consider the detailed formation channels of WD+He star binary systems in our
binary calculations with the {\tt STARS} code.  
To obtain a star as cool as SN~2012Z-S1 at the moment of SN Iax explosion, relatively wide 
WD+He star binary systems with an orbital period of $\gtrsim10\,\rm{days}$ are needed. However, 
such wide WD+He star binary systems cannot be produced by our BPS calculations. Here, the initial 
conditions of the populations and the treatment of the physics of binary evolution
are considered as reasonable as possible in our BPS studies based on current observations and 
theoretical models (e.g., see \citealt{Duch13}). If future 
observations indeed confirm that SN~2012Z had a progenitor system which contained a helium-star 
companion, the improvement of the BPS method may be required. Taking uncertainties on the BPS method 
into account, we conclude that the possibility of a He star donor for SN~2012Z cannot be
excluded, but the expected probability is very low.

Additionally, as mentioned by \citet{McCu14}, the possibility that SN~2012Z-S1 is a single massive 
star that itself exploded cannot be ruled out yet. Fortunately, it may be possible to distinguish 
between the Ch-mass binary model and the single massive-star model through HST imaging when SN 2012Z 
will have faded below the brightness of SN~2012Z-S1 (see also \citealt{McCu14}). The simulations 
for weak deflagration explosions of the Ch-mass WD \citep{Jord12, Krom13, Fink14, Krom15} showed 
that an abundance-enriched bound remnant of the WD will be left after the SN explosion. Also, the 
companion stars \citep{Mari00, Liu12, Liu13b, Liu13a, Liu13c, Pan13} are expected to survive from the SN explosion. If the SN~2012Z produces from 
the C/O WD (or hybrid C/O/Ne WD)+He star scenario, a surviving bound remnant of the WD and He companion star 
probably be detected in the SN remnants though some distinct features during their long-term post-explosion 
evolution such as a spatial velocity (if the binary system can be destroyed by the SN explosion, see \citealt{Jord12}), a particular 
luminosity evolution \citep{Pan13, Fole14} or even a heavy-element enrichment \citep{Liu13c, Pan13}.

\subsection{Uncertainties}
\label{sec:uncertain}

In the SD scenario, only a fairly narrow range in the accretion rate will allow 
stable H- or He-burning to be attained on the surface of the WD, avoiding a nova explosion. 
In this work, the prescription of optically thick wind model from \citet{Hach99} and 
He-retention efficiencies from \citet{Kato04} are used to describe the mass accumulation 
efficiency of accreting WDs. However, strong constraints on the uncertainties of 
the mass-retention efficiencies are still lacking (e.g., \citealt{Shen07, Wolf13, Pier14}).
Different mass-retention efficiencies are expected to somewhat change the results from our binary 
calculations and thus the BPS results (e.g., see \citealt{Bour13, Toon14, Ruit14, Pier14}).
However, we do not expect a significant change in the main conclusions presented in this paper.
This issue will be addressed in detail in a forthcoming study.

In addition, there is a selection effect due to the initial conditions ($M_{\rm{WD}}^{i}$, $M_{2}^{i}$, $\rm{log_{10}}$\,$P^{i}$) 
of the populations. The initial conditions of BPS calculations may sensitively rely 
on the assumed parameters in specific BPS code, which will lead to some uncertainties on 
the BPS results \citep{Toon14}. For instance, CE evolution, current star-formation rate and initial 
mass function. However, current constraints on these parameters (e.g., the CE 
efficiency, see \citealt{Zoro10, De11, Ivan13}) 
are still weak. For a detailed discussion for the effect of different theoretical 
uncertainties, see \citet{Clae14}.

\section{Conclusion and Summary}

\label{sec:summary}

Some potential progenitor scenarios have been proposed to explain the observational properties 
of SNe Iax. Although weak deflagrations of Ch-mass WDs seem to be the most promising scenario 
for SNe Iax \citep{Jord12, Krom13, Fole13, Fink14, McCu14}, including the faint SNe Iax such as 
2008ha-like events \citep{Krom15}. However, the nature of the progenitor systems of SNe Iax is still unclear. 
In this work, we predict observational features of companion stars of binary 
progenitor systems at the moment of the SN Iax explosion by performing detailed 
binary evolution calculations with the 1D stellar evolution code {\tt STARS}, which will be 
very helpful for constraining the progenitor systems of SNe Iax through their pre-explosion 
imaging (e.g., \citealt{McCu14, Fole15}).

Comparing our results with the observations of SN~2012Z-S1, it is found that our detailed binary 
evolution calculations for the C/O WD+He star and hybrid WD+He star Ch-mass channel can contribute significantly 
to the number of systems in the vicinity of SN~2012Z-S1, but it seems the hybrid WD+He star Ch-mass channel
is more favorable for the observations of SN~2012Z-S1. However, our BPS calculations for both the C/O WD+He star and 
the hybrid WD+He star Ch-mass channel produces companion stars as cool as 
the SN~2012Z-S1 only at very low probability. We thus conclude that the possibility of the He donor star as a companion of SN 2012Z is 
low based on our BPS predictions, but the possibility cannot be excluded. A further confirmation needs future 
observations after SN 2012Z will have faded below the brightness of SN~2012Z-S1.

\section*{Acknowledgments}

      We thank the anonymous referee for his/her valuable comments and suggestions that helped us to improve 
      the paper. We would like to thank Elvijs Matrozis for his helpful discussions. This work is supported by the Alexander von Humboldt Foundation. 
      R.J.S. is the recipient of a Sofja Kovalevskaja Award from the Alexander von Humboldt Foundation.
      B.W. acknowledges support from the 973 programme of China (No. 2014CB845700), the NSFC
      (Nos. 11322327, 11103072, 11033008 and 11390374), the Foundation 
      of Yunnan province (No. 2013FB083).

\bibliographystyle{apj}

\bibliography{ref}

\end{document}